\documentclass{article}
\usepackage{amssymb}
\usepackage{bm}
\usepackage[dvips]{graphicx}
\begin{document}

\title{Tunnel magnetoresistance of Fe$_3$O$_4$/MgO/Fe nanostructures}

\author{S.G. Chigarev$^1$, E.M. Epshtein$^1$\thanks{E-mail: epshtein36@mail.ru},
I.V. Malikov$^2$, G.M. Mikhailov$^2$, P.E. Zilberman$^1$\\ \\
$^1$\emph{V.A. Kotelnikov Institute of Radio Engineering and Electronics}\\
    \emph{of the Russian Academy of Sciences, 141190 Fryazino, Russia}\\ \\
    $^2$\emph{Institute of Microelectronics Technology and High Purity Materials} \\
    \emph{of the Russian Academy of Sciences, 142432 Chernogolovka, Russia}}
\date{}
\maketitle

\abstract{A magnetic tunnel junction Fe$_3$O$_4$/MgO/Fe with (001) layer
orientation is considered. The junction magnetic energy is analyzed as a
function of the angle between the layer magnetization vectors under
various magnetic fields. The tunnel magnetoresistance is calculated as a
function of the external magnetic field. In contrast with junctions with
unidirectional anisotropy, a substantially lower magnetic field is
required for the junction switching.} \\ \\

Tunnel magnetic junction is one of the most important objects in spintronics.
The interest to such structures is related with the tunnel
magnetoresistance (TMR) effect used to create magnetic random access memory
(MRAM).

Investigations of magnetic tunnel structures are directed to searching new
effects as well as studying material combinations which are capable to
ensure better characteristics.

Fe$_3$O$_4$/MgO/Fe tunnel junctions seem promising in several extents.
Besides technological advantages, such structures have interesting
physical properties. First, Fe$_3$O$_4$ is so called half metal in which
only the carriers with one spin orientation take part in electric
transport that leads to increasing TMR~\cite{Guilliere}. With this fact as
well as higher resistivity compared to metals, higher spin resistance~\cite{Epshtein}
is related of Fe$_3$O$_4$ layer, which may act as an ideal spin injector.
Second, because of cubic crystallographic symmetry, Fe and Fe$_3$O$_4$
have more than one magnetic anisotropy axes. This increases the number of
stationary configurations of the magnetic junction and, correspondingly,
the number of possible variants of switching between those configurations.
The latter feature is the subject of consideration in present work.

Let us consider a tunnel Fe$_3$O$_4$/MgO/Fe junction with (001) layer
orientation. As mentioned, Fe and Fe$_3$O$_4$ both have the same cubic symmetry but
with different sign of the magnetic anisotropy energy: positive in Fe
and negative (at room temperature) in Fe$_3$O$_4$~\cite{Chikazumi}.
Therefore, there are three easy axes directed along the cube edges [100],
[010], [001] in Fe single crystal, while four easy axes along the cube
diagonals [111], [$\bar1$11], [1$\bar1$1], ]11$\bar1$] in Fe$_3$O$_4$
single crystal. In a thin layer with (001) orientation, the
easy axes lie in the layer plane because of high shape anisotropy (this is
valid when the magnetic anisotropy energy density is low compared to the
demagnetization field energy density $2\pi M^2$, $M$ being the saturation
magnetization). In such a case, the easy axes in Fe(001) layer will be [100]
and [010], while those in Fe$_3$O$_4$(001) layer will be [110] and
[$\bar1$10], whereas [100] and [010] axes will be hard ones.

The anisotropy energy density in Fe is higher by several times (in magnitude) than that
in Fe$_3$O$_4$ (in single crystals at room temperature, $4.7\times10^4$
J/m$^3$ and $-1.2\times10^4$ J/m$^3$, respectively~\cite{Chikazumi}).
Therefore, Fe$_3$O$_4$ layer will be switched earlier than Fe one under
external magnetic field, other things being equal.

The presence of the MgO barrier layer avoids exchange coupling between Fe and
Fe$_3$O$_4$ layers. As to the dipole magnetic interaction between the
layers, such a coupling is a weak edge effect when the layer thickness is
small compared to the layer lateral sizes. So we assume the responses of
the Fe and Fe$_3$O$_4$ layers to magnetic field to be mutually
independent.

Let us consider behavior of the Fe$_3$O$_4$(001) layer with [100] (hard)
axis parallel to the Fe layer [100] (easy) axis. The external magnetic
field is assumed to be directed along the same axis. Let us track how the Fe$_3$O$_4$
layer magnetization direction changes under varying the applied magnetic
field $H$.

The magnetic energy density takes the form
\begin{equation}\label{1}
  U(\theta)=-MH\cos\theta-|K|\cos^2\theta\sin^2\theta,
\end{equation}
where $K$ is the (negative) anisotropy energy density, $\theta$ is the
angle between the layer magnetization vector and [100] axis.

The equilibrium condition is the equality
\begin{equation}\label{2}
  \frac{dU}{d\theta}=0,
\end{equation}
the equilibrium stability condition is the inequality
\begin{equation}\label{3}
  \frac{d^2U}{d\theta^2}>0.
\end{equation}
The condition~(\ref{2}) with Eq.~(\ref{1}) taking into account is reduced
to a trigonometric equation
\begin{equation}\label{4}
  \left(2\cos^3\theta-\cos\theta-h\right)\sin\theta=0,
\end{equation}
where
\begin{equation}\label{5}
  h=\frac{MH}{2|K|}=\frac{H}{H_a}
\end{equation}
is the dimensionless magnetic field, $H_a=2|K|/M$ is the anisotropy field.

In the angle interval from 0 to $\pi$, Eq.~(\ref{4}) has the following
solutions stable in different ranges of $h$ values:
\begin{equation}\label{6}
  \theta_1=0,
\end{equation}
stability at $h>1$;
\begin{equation}\label{7}
  \theta_2=\pi,
\end{equation}
stability at $h<-1$;
\begin{equation}\label{8}
  \theta_3=\arccos\left(\sqrt\frac{2}{3}\cos\left(\frac{1}{3}\arccos\frac{h}{h_0}\right)\right)
\end{equation}
and
\begin{equation}\label{9}
  \theta_4=\arccos\left(-\sqrt\frac{2}{3}\cos\left(\frac{\pi}{3}-\frac{1}{3}
  \arccos\frac{h}{h_0}\right)\right),
\end{equation}
where $h_0=\sqrt{\displaystyle\frac{2}{27}}\approx0.272$, stability at
$|h|<h_0$;
\begin{equation}\label{10}
    \theta_5=\arccos\left(-\sqrt\frac{2}{3}\cos\left(\frac{\pi}{3}+\frac{1}{3}
    \arccos\frac{h}{h_0}\right)\right),
\end{equation}
instability;
\begin{equation}\label{11}
  \theta_6=\arccos\left(\sqrt\frac{2}{3}\cosh\left(\frac{1}{3}{\rm arcosh}
  \frac{h}{h_0}\right)\right),
\end{equation}
stability at $h_0<h<1$;
\begin{equation}\label{12}
  \theta_7=\arccos\left(-\sqrt\frac{2}{3}\cosh\left(\frac{1}{3}{\rm arcosh}
  \left(-\frac{h}{h_0}\right)\right)\right),
\end{equation}
stability at $-1<h<h_0$.

\begin{figure}
\includegraphics{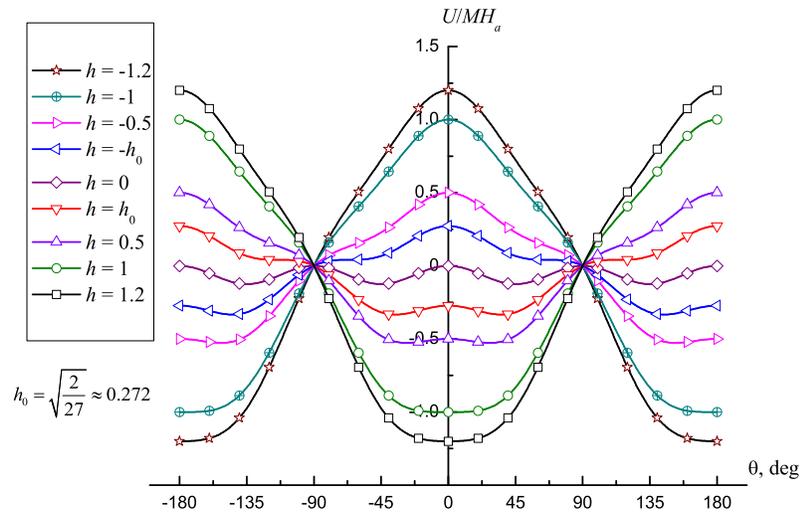}
\caption{The dependence of the (dimensionless) magnetic energy of the
Fe$_3$O$_4$ layer on the magnetization vector orientation under various
magnetic fields.}\label{fig1}
\end{figure}

It is seen that two solutions, $\theta_3$ and $\theta_4$, are stable in the
$-h_0<h<h_0$ range. Realization of either depends on the prehistory. With
decreasing the magnetic field from $h=+\infty$ to certain value $h<-1$ at
which the Fe layer magnetization does not reverse yet, the following
sequence of states takes place:
\begin{equation}\label{13}
  \theta_1\stackrel{1}{\longrightarrow}\theta_6\stackrel{h_0}{\longrightarrow}\theta_3
  \stackrel{-h_0}{\longrightarrow}\theta_7\stackrel{-1}{\longrightarrow}\theta_2;
\end{equation}
the indices over the arrows show the magnetic field values at which
corresponding switching occurs. With changing the field in the opposite
direction another sequence takes place:
\begin{equation}\label{14}
  \theta_2\stackrel{-1}{\longrightarrow}\theta_7\stackrel{-h_0}{\longrightarrow}\theta_4
  \stackrel{h_0}{\longrightarrow}\theta_6\stackrel{1}{\longrightarrow}\theta_1.
\end{equation}
These sequences may be tracked also in Fig.~\ref{fig1} where the (dimensionless)
magnetic energy is shown as a function of $\theta$ angle with various $h$
values.

\begin{figure}
\includegraphics{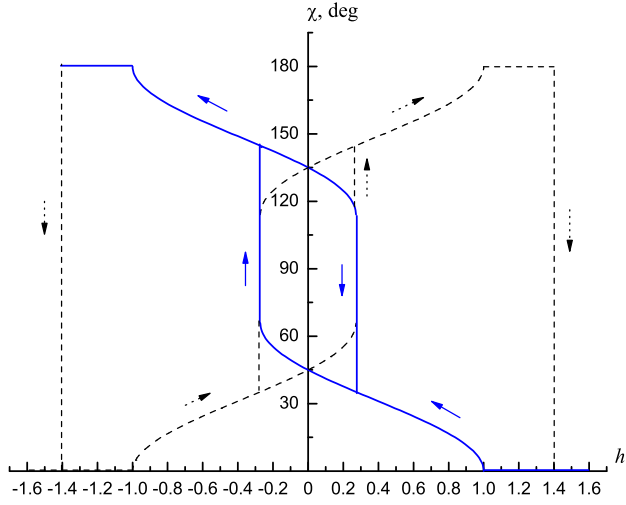}
\caption{The field dependence of the $\chi$ angle between the Fe and
Fe$_3$O$_4$ layer magnetization vectors before (solid lines and arrows)
and after (dashed lines and dotted arrows) switching the Fe layer which has
higher anisotropy energy.}\label{fig2}
\end{figure}

In Fig.~\ref{fig2}, $\chi$ angle between the layer magnetization vectors is shown
as a function of the magnetic field before (solid lines and arrows) and
after (dashed lines and dotted arrows) switching the Fe layer which has
higher anisotropy energy, so that higher magnetic field is required for
its switching. This layer, magnetized along the positive direction of [100]
axis initially, is switched to the opposite direction by some magnetic
field $h=h_1<-1$ having negative direction. The sequence~(\ref{14}) does
not realize in this case.

The change of the junction magnetic configuration manifests itself as
change of the junction resistance. The conductance of a magnetic tunnel
junction with $\chi$ angle between the layer magnetization vectors takes
the form~\cite{Utsumi}
\begin{equation}\label{15}
  G(\chi)=G_P\cos^2\frac{\chi}{2}+G_{AP}\sin^2\frac{\chi}{2},
\end{equation}
where $G_P,\,G_{AP}$ are the junction conductances under parallel ($\chi=0$)
and antiparallel ($\chi=\pi$) relative orientation of the layers,
respectively.

It is convenient to take the following ratio as a measure of the junction
resistance change:
\begin{equation}\label{16}
  F(h)\equiv\frac{R(h)-R_P}{R_P}=\frac{\rho(1-\cos\chi(h))}{2+\rho(1+\cos\chi(h))},
\end{equation}
where $R(h)=1/G(\chi(h))$, $\rho=(R_{AP}-R_P)/R_P$ is TMR defined in a
usual way~\cite{Guilliere}. The latter is related with the layer spin
polarizations $P_1,\,P_2$~\cite{Guilliere}:
\begin{equation}\label{17}
  \rho=\frac{2P_1P_2}{1-P_1P_2}.
\end{equation}
With~$P_1=0.44$ (Fe~\cite{Moodera}), $P_2=1$ (Fe$_3$O$_4$) we have
$\rho\approx1.6$.

\begin{figure}
\includegraphics{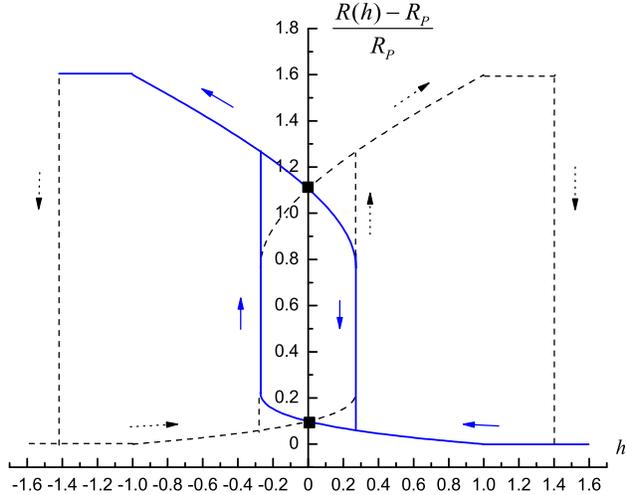}
\caption{The junction resistance change as a function of the magnetic
field. The black squares mark the stationary states with different
resistances between which the switching occurs.
}\label{fig3}
\end{figure}

To obtain the junction resistance change as a function of the magnetic field
$F(h)$, the $\chi(h)$ dependence should be substitute to Eq.~(\ref{16}).
With the minimum magnetic energy analysis made above taking into account, we obtain the
results shown in Fig.~\ref{fig3}. As in Fig.~\ref{fig2}, the solid and dashed lines show the
resistance change before and after the Fe layer switching, respectively. A
possibility is seen of the switching between stationary states (marked
with black squares) with different resistances.

In comparison with the standard TMR, where the magnetic field equal to the
anisotropy field of the layer is required for switching, the substantially lower field
$h=h_0$ is needed in the considered case.

The work was supported by the Russian Foundation for Basic Research,
Grants Nos.~08-07-00290 and~10-07-00160.


\begin{thebibliography}{1}
\bibitem{Guilliere}
M. Guilliere, Phys. Lett. A \textbf{54}, 225 (1975)
\bibitem{Epshtein}
E.M. Epshtein, Yu.V. Gulyaev, P.E. Zilberman, J. Magn. Magn. Mater.
\textbf{312}, 200 (2007)
\bibitem{Chikazumi}
S. Chikazumi. Physics of Magnetism (Clarendon, Oxford, 1997)
\bibitem{Utsumi}
Y. Utsumi, Y. Shimizu, H. Miyazaki, J. Phys. Soc. Japan \textbf{68}, 3444 (1999)
\bibitem{Moodera}
J.S. Moodera, G. Mathon, J. Magn. Magn. Mater. \textbf{200}, 248 (1999)

\end{thebibliography}
\end{document}